\documentclass[twocolumn,showpacs,amsmath,amssymb,prb]{revtex4}
\usepackage{graphicx}
\usepackage{color}
\usepackage{amsmath}
\usepackage{amssymb}
\def \be {\begin{equation}}
\def \ee {\end{equation}}

\begin{document}

\newcommand{\FM}[1]{\textcolor{red}{\bf [#1]}}
\newcommand{\AS}[1]{\textcolor{blue}{\bf [#1]}}
\preprint{1}

\title{Self-consistent calculation of the electron distribution near
a Quantum-Point Contact in the integer Quantum Hall Effect}
\author{A. Siddiki}%
\author{F. Marquardt}
\affiliation{Physics Department, Arnold Sommerfeld Center for
Theoretical Physics, and
Center for NanoScience, \\
Ludwig-Maximilans-Universit\"at, Theresienstrasse 37, 80333
Munich, Germany}
\date{\today}
\begin{abstract}
In this work we implement the self-consistent Thomas-Fermi-Poisson
approach to a homogeneous two dimensional electron system (2DES).
We compute the electrostatic potential produced inside a
semiconductor structure by a quantum-point-contact (QPC) placed at
the surface of the semiconductor and biased with appropriate
voltages. The model is based on a semi-analytical solution of the
Laplace equation. Starting from the calculated confining potential,
the self-consistent (screened) potential and the electron
densities are calculated for finite temperature and magnetic
field. We observe that there are mainly three characteristic
rearrangements of the incompressible "edge" states, which
will determine the current distribution near a QPC.
\end{abstract}

\maketitle
\section{\label{sec:1} Introduction}

A quantum point contact (QPC) is constructed by geometric or
electrostatic confinement of a two dimensional electron system
(2DES). The conductance through them is
quantized~\cite{Wees88:848,Whraham88:209} and they play a crucial
role in the field of mesoscopic quantum transport. Their
properties have been investigated in a wide variety of
experiments, which include the observation of the 0.7
anomaly~\cite{Thomas96:135,Kristensen00:10950}, quantum dots
coupled to QPCs~\cite{Ludwig06:377}, Quantum-Hall effect (QHE)
based Mach-Zender~\cite{Heiblum03:415,Neder06:016804} and
Aharonov-Bohm interferometers. This has lead to extensive
investigations of the electrostatic and transport properties of
QPCs, both with and without a quantizing magnetic field. Many
different techniques have been used to find the electronic density
distribution near a QPC, ranging from numerical
Poisson--Schr\"odinger solutions~\cite{Macucci02:39} to
spin-density-functional theory~\cite{Hirose03:026804} and
phenomenological approaches~\cite{Reilly06:135}. It has been
possible to treat realistic samples mostly only within simplified
electrostatic calculations, neglecting screening effects. On the
other hand, when including interactions the calculations become
more complicated, thus one usually sacrifices handling realistic
geometries.

Recent experiments have succeeded in developing and analyzing a
QHE based electronic Mach-Zender interferometer
(MZI)~\cite{Heiblum03:415}, making use of the integer QHE edge
states~\cite{Neder06:016804} as single-channel chiral quantum
wires. A key element of these experiments are the QPCs, which play
the role of the beam splitters of the optical setup. The
electrostatic potential and electronic density distributions in
and near the QPCs play an important role in understanding the
rearrangement of the edge states involved. Moreover, the
electron-electron interaction has been
proposed~\cite{Neder06:016804} as one of the origins of dephasing
in such an electronic MZI, such that a self-consistent calculation
of the electrostatic potential may also be viewed as a first step
towards a quantitative understanding of this issue. So far, the
theoretical description of dephasing in the electronic MZI via
classical~\cite{Seeling01:245313,Florian04:056805,Florian04:125305,Forster05:}
or quantum noise fields~\cite{Florian05:788,Florian06:condmat1}
and other approaches~\cite{Pilgram05:condmat} has focused on
features supposed to be independent of its specific realization
(see Ref.~[\onlinecite{Florian06:condmat}] for a recent review).
However, a more detailed analysis of the QHE related physics,
taking account of interaction effects, will certainly be needed
for a direct comparison with experimental data. In this paper, we
will provide a detailed numerical analysis of the electrostatics
of QPCs in the integer QHE, assuming geometries adapted to those
used in the MZI experiment. Our work will produce the electron
density and electrostatic potential, based on the self-consistent
Thomas-Fermi-Poisson approximation, to which we refer as TFA in
the following.

We would like to point out the following observation regarding the
Mach-Zehnder experiment, where a yet-unexplained beating pattern
observed in the visibility (interference contrast) as a function
of bias voltage was surprisingly found to have a period
independent of the length of the interferometer arms. Such a
result would seem less surprising if all the relevant interaction
physics leading to the beating pattern were actually taking place
in the vicinity of the QPC. This provides strong encouragement for
future more detailed work on the coherent transport properties of
these QPCs.

Although it has been more than two decades since the discovery of
the quantized Hall effect~\cite{vKlitzing80:494}, the microscopic
picture of current distribution in the sample and the interplay of
the current distribution with the Hall plateaus is still under
debate. In recent experiments, the Hall potential distribution and
the local electronic compressibility have been investigated in a
Hall bar geometry by a low-temperature scanning force
microscope~\cite{Ahlswede02:165} and by a
single-electron-transistor~\cite{Yacoby00:3133}, respectively.
This has motivated theoretical~\cite{Guven03:115327} work, where a
self-consistent TFA calculation has been used to obtain
electrostatic quantities.

Self-consistent screening calculations show that the 2DES contains
two different kinds of regions, namely the quasi-metallic
compressible and quasi-insulating incompressible
regions~\cite{Chklovskii92:4026,Siddiki03:125315}. The electron
distribution within the Hall bar depends on the "pinning" of the
Fermi level to highly degenerate Landau levels. Wherever the Fermi
level lies within a Landau level with its high density of states
(DOS), the system is known to be compressible (leading to
screening and correspondingly to a flat potential profile),
otherwise it is incompressible, with a constant electron density
and, in general, a spatially varying potential due to the absence
of screening. Moreover, based on these results for the potential
and density distributions, one may employ a local version of Ohm's
law  (together with Maxwell's equations and an appropriate model
for the conductivity tensor) to calculate the current
distribution, imposing a given overall external current for the
in-plane geometry . These results are mostly consistent with
experiments except that within the self-consistent TFA one obtains
an incompressible strip (IS) for a large interval of magnetic
field values which leads to coexistence of several IS's with
different local filling factors. Recently, this theory has been
improved in two aspects~\cite{siddiki2004,Siddiki04:condmat}: (i)
the finite extent of the wave functions was taken into account in
obtaining electrostatic quantities (rather than using delta
functions), (ii) the findings of the full Hartree calculations
were simulated by a simple averaging of the local conductivities
over the Fermi wave length , thereby relaxing the strict locality
of Ohm's law for realistic sample sizes.  A very important outcome
of this model is that there can exist only one incompressible edge
state at one side of the sample for a given magnetic field value.
Indeed this is differing drastically from the
Chklovskii-Shklovskii-Glazman (CSG)~\cite{Chklovskii92:4026} and
the Landauer-B\"uttiker~\cite{Buettiker88:317} picture, where more
than one edge state can exist and is necessary to "explain" the
QHE. In the CSG scheme a non-self-consistent TFA (which is called
the "electrostatic approximation") was used. However, it is clear
that if the widths of the IS's (where the potential variation is
observed) become comparable with the magnetic length, the TFA is
not valid, thus the results obtained within this model are not
reliable any more. In principle, similar results to
Ref.~[\onlinecite{siddiki2004}] were reported by T. Suzuki and T.
Ando~\cite{Suzuki93:2986}, quite some time ago and recently by S.
Ihnatsenka and I. V. Zozoulenko~\cite{Igor06:155314} in the
context of spin-density-functional theory. With the improvements
on the self-consistent TFA mentioned above, together with taking
into account the disorder potential~\cite{Siddiki:ijmp} and using
the self-consistent Born approximation~\cite{Ando82:437} to
calculate the local conductivity tensor, one obtains well
developed Hall plateaus, with the longitudinal resistivity
vanishing to a very high accuracy, and one is also able to
represent correctly the inter-plateau transition regions. Wherever
one observes an IS, the longitudinal conductivity becomes zero,
and as a consequence also the corresponding local resistance (and
the total resistance) vanishes. Thus, according to Ohm's law, the
current flows through the incompressible region. In addition, the
Hall conductance becomes equal to the local value of the quantized
conductance. Finally all the three experimentally
observed~\cite{Ahlswede01:562} qualitatively different regimes of
how the Hall potential drops across the sample have been
reproduced theoretically without artifacts of the
TFA~\cite{Guven03:115327}. The model described above has also been
successfully applied to an electron-electron bilayer
system~\cite{Bilayersiddiki06:} and provided a qualitative
explanation~\cite{siddikikraus:05} of the magneto-resistance
hysteresis that has been reported recently~\cite{tutuc,pan}. For
all of these reasons, we feel confident in applying this theory to
our analysis of the MZI setup.

Motivated by the experimental and theoretical findings
ascertaining the importance of the interaction effects in the
integer Quantum Hall regime, in this work we will show that the
mutual Coulomb interaction between the electrons leads to
interesting non-linear phenomena in the potential and electron
distribution in close proximity of the QPCs. Based on the
self-consistent Thomas-Fermi-Poisson approximation (TFA), we will
consider realistic QPC geometries and examine the distribution of
the incompressible regions depending on the field strength and
sample parameters.

The rest of this paper is organized as follows: In
Sec.~\ref{sec:electroststics} the electrostatic potential produced
by an arbitrary surface gate will be discussed, by solving the
Laplace equation without screening effects. In Sec.~\ref{sec:TFA}
we review the TFA in a 2DES. In Sec.~\ref{sec:numerics} we will
first present the well known general results of the TFA for a
homogeneous 2DES at zero magnetic field $B$ and zero temperature,
and we will investigate the electron density and electrostatic
potential profiles of a (i) simple square gate geometry and (ii) a
generic QPC, before (iii) systematically investigating the
positions of the incompressible strips depending on magnetic field
and geometric parameters. We conclude with a discussion in
Sec.~\ref{sec:summary}

\section{\label{sec:electroststics}Electrostatics of the gates}

As mentioned in the introduction, there is a tradeoff between
simulating realistic QPC geometries and including the interaction
effects within a reasonable approximation. In this paper, we
present an intermediate approach, which considers realistic QPC
structures but interactions of the electrons are handled within a
Thomas-Fermi approximation (TFA), which is valid for relatively
"large" QPCs ($\gtrsim 100$ nm). One can obtain, in a
semi-analytical fashion, the electrostatic potential generated by
an arbitrary metallic gate at the surface by solving the Laplace
equation for the given boundary conditions. Afterwards, it is
possible to obtain the electron and potential distribution in the
2DES, within the TFA, both for vanishing and finite magnetic
fields ($B$), and at low temperatures at $B>0$.

Here we briefly summarize the semi-analytical model developed by
J.Davies and co-workers\cite{Davies95:4504}. The aim of this
section is to calculate the electrostatic potential on a plane at
some position $z$ below the surface of the semiconductor, which is
partially covered by a patterned gate. The surface occupies the
$z=0$ plane and z is measured into the material. The un-patterned
surface is taken to be pinned so we can set the potential
$V_{up}({\bf{r}},0)=0$ there, with $V_{gate}({\bf{r}},0)=V_{g}$ on
the gate. We use lower-case letters like ${\bf{r}}= (x,y)$ to
denote two-dimensional vectors with the corresponding upper-case
letters for three dimensional vectors like ${\bf R}= (x,y,z) =
({\bf r},z)$. Thus the problem is to find a solution,
$V_{ext}({\bf{R}})$, to the Laplace equation
$\nabla^{2}V_{ext}=0$, given the value on the plane $z= 0$, and
subject to the further boundary condition $\partial
V_{ext}/\partial z\rightarrow0$ as $z\rightarrow\infty$.
\begin{figure}
{\centering \includegraphics[width=0.7\linewidth]{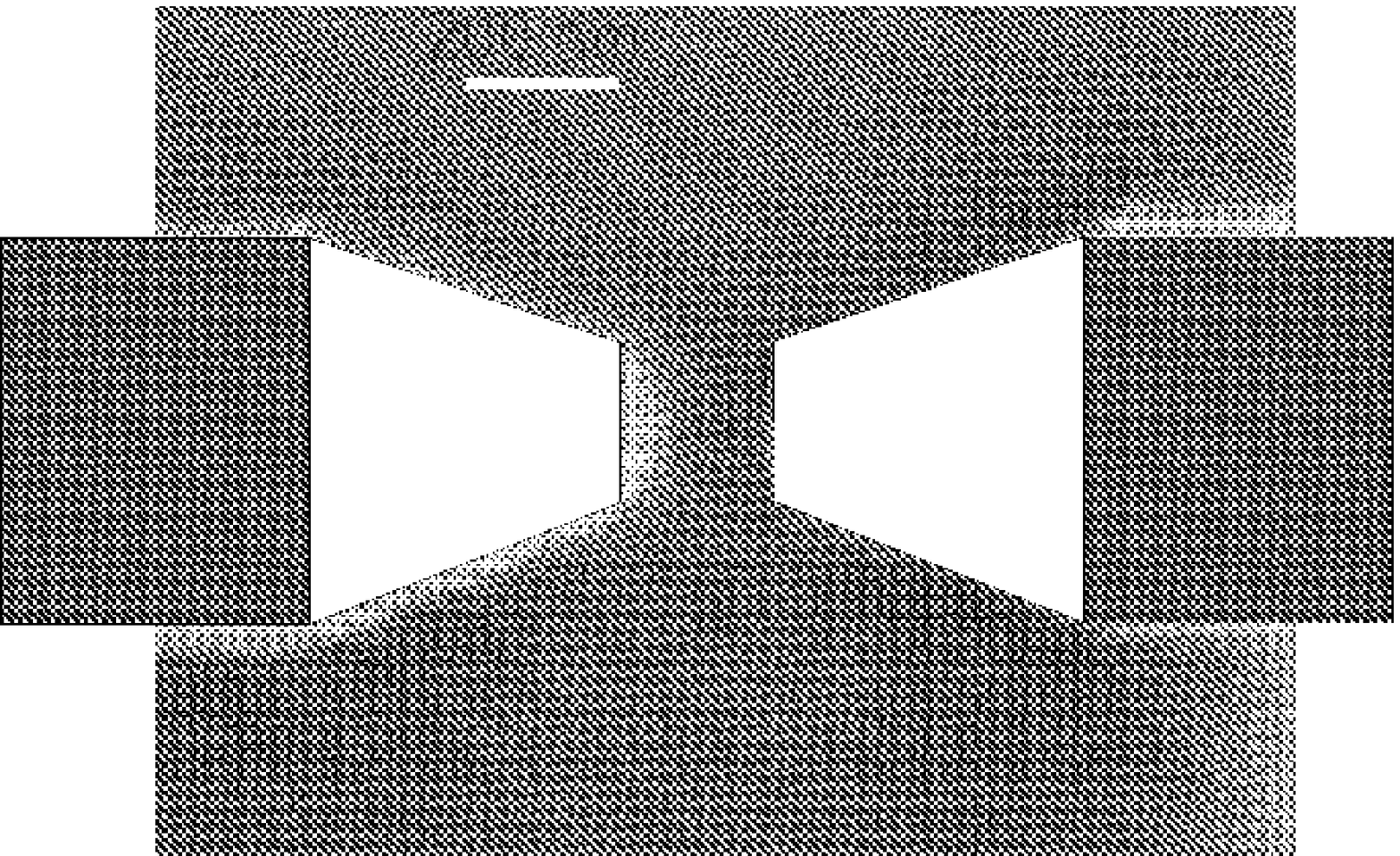}}
\vspace{1cm} \caption{ \label{fig:qpcgeo} The image of the QPC
(gray scale). The polygons are used to define the gates on the
two-dimensional mesh. The 2DES resides under the dark (grey)
regions, with a bulk electron density of
$1.7\times10^{-11}cm^{-2}$ (see Ref.~[\onlinecite{Heiblum03:415}],
due to memory limitations the quality of the figure is reduced.
The white line scales $200$nm). }
\end{figure}
One route is to start by making a two-dimensional Fourier
transform from $V_{ext}({\bf{r}},0)$ to
$\tilde{V_{ext}}({\bf{q}},0)$. The dependence on z is a decaying
exponential to satisfy Laplace's equation and the boundary
condition at $z=\infty$:
$\tilde{V}_{ext}({\bf{q}},z)=\tilde{V}_{ext}({\bf{q}},0)\exp(-|qz|)$.
This multiplication of the Fourier transform is equivalent to a
convolution in real space. Taking the two-dimensional inverse
Fourier transform of $\exp(-|qz|)$ leads to the general result.
\be \label{eq:davies}V_{\rm{ext}}({\bf{r}},z)=\frac{1}{\kappa}
\int{\frac{|z|}{2\pi(z^{2}+|{\bf{r}}-{\bf{r'}}|^{2})^{3/2}}}V_{gate}({\bf{r'}},0)d{\bf{r'}},
\ee where $\kappa$ is the dielectric constant of the considered
hetero-structure. Now one can evaluate the potential in the plane
of the 2DES, $z=d$, for a given gate and potential distribution on
the surface. The derivation of some important shapes like
triangle, rectangle and polygons is provided in the work cited
above, which has been successfully applied to quantum dot
systems\cite{Stefan}. For our geometry, we will use the result for
the polygons.

\section{\label{sec:TFA}Electron-electron interaction: Thomas-Fermi-Poisson Approximation}
The main assumption of this approximation is that the external
(confining) potential varies smoothly on the length scale of the
magnetic length, $l_{b}=\sqrt{\hbar /(m \omega_c)}$, where $m$ is
the effective mass of an electron in a GaAs/AlGaAs
hetero-structure, and $\omega_c$ is the cyclotron frequency given
by $\omega_c=eB/m$ for the magnetic field strength $B$. At the
magnetic field strengths of our interest, where the average
filling factor ($\bar{\nu}$) is around 2, i.e. $B>5 T$, $l_{b}$ is
on the order of 10 nanometers, hence the TFA is valid. We note
that spin degeneracy will not be resolved in our calculations.
This can be done if the cyclotron energy is much larger than the
Zeeman energy (i.e. effectively we set $g=0$).

In the following, we briefly summarize the self-consistent
numerical scheme adopted in this work.  We will assume the 2DES to
be located in the plane $z=85$nm with a (surface)  number density
$n_{\rm el}(x,y)$. We consider a rectangle of finite extent $a_x
\times a_y$ in the $xy$-plane, with periodic boundary conditions.
The (Hartree) contribution $V_H(x,y)$ to the potential energy of
an electron caused by the total charge density of the 2DES can be
written as \cite{Oh97:13519}
\be  \label{hartree} V_H(x,y)= \frac{2e^2}{\bar{\kappa}}
\int_{0}^{a_{x}}\int_{0}^{a_{y}} \!dx' \!dy' K(x,x',y,y')\,n_{\rm
  el}(x',y'),
\ee
where $-e$ is the electron charge, $\bar{\kappa}$ an average
background dielectric constant, \cite{Oh97:13519} and the kernel
$K(x,x',y,y')$ describes the solution of Poisson's equation with
appropriate boundary conditions. This kernel can be found in a well known
text book \cite{Morse-Feshbach53:1240}. The electron density in turn is calculated in the
Thomas-Fermi  approximation (TFA) \cite{Oh97:13519}
\be \label{thomas-fermi}
 n_{\rm el}(x,y)=\int dE\,D(E)f\big( [E+V(x,y)-\mu^{\star}]/k_{B}T \big),
\ee
with $D(E)$ the relevant (single-particle) density of states
(DOS), $f(s)=[1+e^s]^{-1}$ the Fermi function, and $\mu^{\star}$
the electrochemical potential. The total potential energy of an
electron, $V(x,y)=V_{\rm ext}(x,y)+V_H(x,y)$, differs from
$V_H(x,y)$ by the contribution due to external potentials, e.g.
the confinement potential generated by the QPC (see figure
\ref{fig:barepot}), potentials due to the donors etc. The local
(but nonlinear) TFA is much simpler than the corresponding quantum
mechanical calculation and yields similar results if $V(x,y)$
varies slowly in space\cite{siddiki2004}, i.e. on a length scale
much larger than typical quantum lengths such as the extent of
wave functions or the Fermi wavelength.

\section{\label{sec:numerics}Numerical calculations} The equations (\ref{hartree})
and (\ref{thomas-fermi}) have to be solved self-consistently for a
given temperature and magnetic field, until
convergence is obtained. In our scheme we start with vanishing
field and at zero temperature to obtain the electrostatic
quantities and use these results as an initial value for the
finite temperature and field calculations. For $B,T>0$ we start
with a relatively high temperature and reduce $T$ stepwise in
order to obtain a good numerical convergence.

\subsection{Zero magnetic field}
In this section we review the theory of
screening in a homogeneous 2DES.

Mesoscopic systems like quantum dots, Hall bars, or
any edges of quasi-2D electron systems are defined by lateral
confinement conditions, which lead to an inhomogeneous electron
density. An exact treatment of the mutual interactions of the
electrons in such systems is only possible for quantum dots with
very few (less than ten) electrons.

The total potential seen by any electron is given by the sum of
the external potential (describing the confinement) and the
Hartree potential given by Eq. (\ref{hartree}), where the electron
density in turn is determined self-consistently by the effective
single-particle potential $V_{ext} + V_H$.

Now consider a 2DES in the $xy$- plane (with vanishing thickness)
and having the charge density \be
n_{el}^{3D}(\vec{r})=n_{el}^{2D}({\bf r})\delta(z)=\int
\frac{d^2q}{(2\pi)^2}n^q e^{i{\bf q r}}\delta(z) \ee with ${\bf
q}=2\pi(n_x/a,n_y/b)$, where $n^q$ is the $q$- component of the
Fourier transformed electron density. We want to obtain the
effects of an external perturbation $\delta V_{ext}({\bf r}, z)$,
whose Fourier components in the plane $z = 0$ are $\delta
V_{ext}^q(0)$. This potential induces a charge density $\delta
n^q$, which in turn leads to an induced potential \be \delta
V_{ind}^{q}(z)=\frac{2\pi e^2}{\kappa q}e^{-q|z|}\delta n^q,\ee
that has the tendency to screen the applied external potential.
Within the TFA, the induced density is related to the overall
screened potential $V_{scr}$ by $ \delta n^q=-D_{T}^{2D}\delta
V_{scr}^q(0)$, where $D_{T}^{2D}$ is the thermodynamic DOS defined
as $D_T=\int dE D(E)\frac{df}{d \mu}$.

Employing $\delta V_{scr}=\delta V_{ext}+\delta V_{ind}$, this
yields \be \delta V_{scr}^q(0)=\frac{\delta
V^{q}_{ext}(0)}{\varepsilon({\bf q})}, \ee where \be
\varepsilon({\bf q})=1+\frac{q_{TF}}{q}\ee is the 2D dielectric
function with the Thomas-Fermi momentum \be q_{TF}=\frac{2\pi
e^2}{\kappa}D_T^{2D}. \ee Then the self-consistent potential at
distance $|z|$ from the 2DES is \be \delta V_{scr}^q(z)=\delta
V_{ext}^q(z)-\frac{q_{TF}}{q+q_{TF}}e^{-q|z|}\delta V_{ext}^q(0)
,\ee i.e., the screening effect of the 2DES decreases
exponentially with $|z|$.

In the limit $B=0$, $T\rightarrow 0$ and
with $E_F=\mu^{\star}(B=0,T=0)$, Eq.~(\ref{thomas-fermi}) reduces
to
\be \label{linear-resp} n_{\rm el}(x,y)=D_0\,\big(E_F-V(x,y)\big)
\, \theta\big(E_F-V(x,y)\big) , \ee
where $D_0$ is the constant DOS for a 2DES given by $D_0=m/(\pi
\hbar^2)$. This is a linear relation between $V(x,y)$ and $n_{\rm
el}(x,y)$ for all $V(x,y)< E_F$.

Now we apply these results to determine the screening of a given
periodic charge distribution in the plane $z=0$, which creates an
external potential $V_{ext}({\bf r},0)=\sum_q V^qe^{i{\bf q r}}$
in this plane. The self-consistent potential in a 2DES then is
described by: \be \label{eq:qdep} V_{scr}(\textbf{r},z)=\sum_q
V_{scr}^q(z)e^{i{\bf q r}}, \quad \quad V_{scr}^q(z)=V^q e^{-q
z}\big( 1+\frac{2}{qa_{\rm B}^{\star}} \big)^{-1}. \ee The
dielectric function $\epsilon (q)$ can be expressed in terms of
the effective Bohr radius $a_{\rm B}^{\star}=\bar{\kappa} \hbar^2
/(m e^2)$ (for GaAs $a_{\rm B}^{\star} =9.8\,$nm),
since\cite{Stern67:546,Wulf88:4218} $2/a_{\rm B}^{\star}=2 \pi e^2
D_0/\bar{\kappa}$, with $q=2\pi/a$. We will assume that
$\varepsilon (q)  \gg 1$, so that the TFA is valid for $B\gtrsim
1\,$T, i.e. $l_m \lesssim 30\,$nm. We also note that the $q=0$
component is cancelled by the homogeneous donor distribution,
assuring overall charge neutrality.

\subsection{Simple example: Square gate barrier}
\begin{figure}
{\centering \includegraphics[width=1.\linewidth]{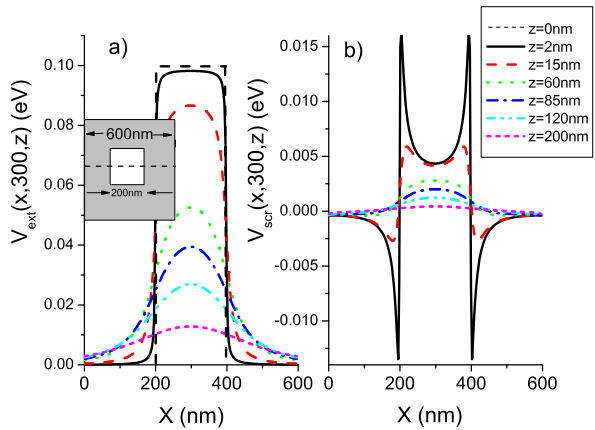}
}
\caption{ \label{fig:sqrpot} External potential (left panel) seen
by a 2DES at different distances z and the corresponding screened
potentials (right panel). The separating dielectric material is
assumed to be GaAs with $\kappa=12.4$ and the calculations are
done at $T=0$K.}
\end{figure}

We start our discussion by a simple example that presents the
features of non-linear screening in a 2DES. We assume a negatively
charged metallic square gate depicted by the white area in the
inset of \ref{fig:sqrpot}a, located at the center of a cell that
is periodically continued throughout the plane (with periods
$a_x=a_y=600$nm). The square is of size $200$nm, and it is kept at
the gate potential, $V_{gate}=-0.1$V. In Fig.\ref{fig:sqrpot} we
show the external and the screened potential for different
separation distances of the 2DES and the gate, calculated along
the dashed line shown in the inset, in the plane of the 2DES.

In the left panel, the external potential has been plotted, with
the dashed line representing the barrier (gate potential) on the
surface. We observe that the potential profile becomes smooth
quickly due to the exponential decay of the amplitude of Fourier
components at large $q$ with increasing $z$\cite{Siddiki:Diss}.

In contrast, the screened potential displays an interesting,
strong feature close to the edges of the gate ($x\sim 200$ and
$x\sim 400$), when the separation distance is relatively small
($z<60$nm). This is nothing but the manifestation of the $q$
dependent screening given in Eq.(~\ref{eq:qdep}): The large $q$
components of the potential remain (almost) unaffected by
screening, whereas the low $q$ (long wavelength) components are
well screened. As a result, we observe sharp peaks near the edge
of the gate for small distances $z$, which turn into "shoulders"
at larger $z$. We should caution, however, that for $z<60$nm the
validity of the TFA may become questionable, since the potential
then changes rapidly on the scale of the Fermi wavelength.

This simple example already demonstrates the strongly non-linear
behavior of the screening, which can be summarized as follows: (i)
the strongly varying part (high $q$ components) of the external
potential remains (almost) unscreened by the 2DES, but its
amplitude decreases fast with increasing separation $z$, whereas
(ii) the slowly varying part (small $q$ components) is well
screened by the 2DES, but its amplitude decays much slower for
large separation distances. Indeed this non-linearity
($q$-dependence of $\varepsilon$) leads to peculiar effects both
on electrostatics and transport properties of the QPCs, depending
on the geometry and the structure of the sample. In the next
section we will look for such effects with regard to the QPCs.

\subsection{Simulation of the QPC}
In this section we will first obtain the \emph{bare confinement
potential} created by the QPC for the geometry given in figure
\ref{fig:qpcgeo}, and then go on to discuss the effects of
screening. The potential generated by such gates can be calculated
by the scheme proposed by Davies \emph{et
al.}\cite{Davies95:4504}.

The model parameters are taken from the relevant experimental
samples\cite{Neder06:016804,Neder06:priv}, where the applied gate
voltage is $-0.3$V, the width at the tip is about $200$nm, and the
tip separation $\Delta y\sim300$nm. The 2DES is taken to be $85$nm
below the surface.

\begin{figure}
{\centering \includegraphics[width=1.\linewidth]{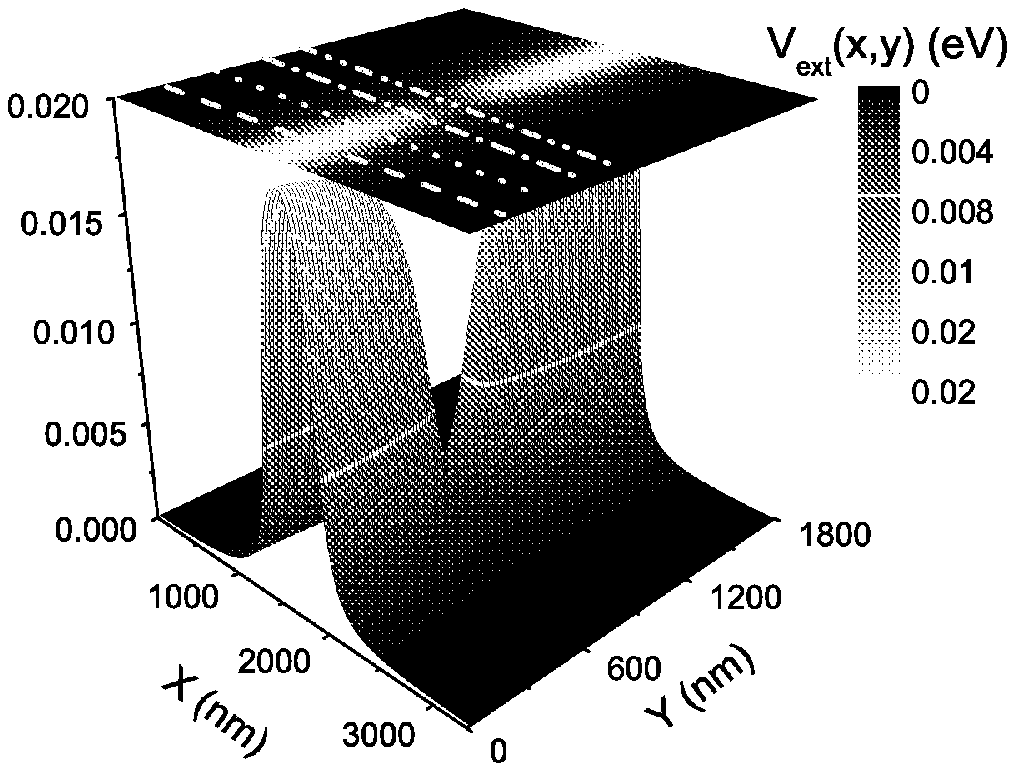}
\includegraphics[width=1.\linewidth]{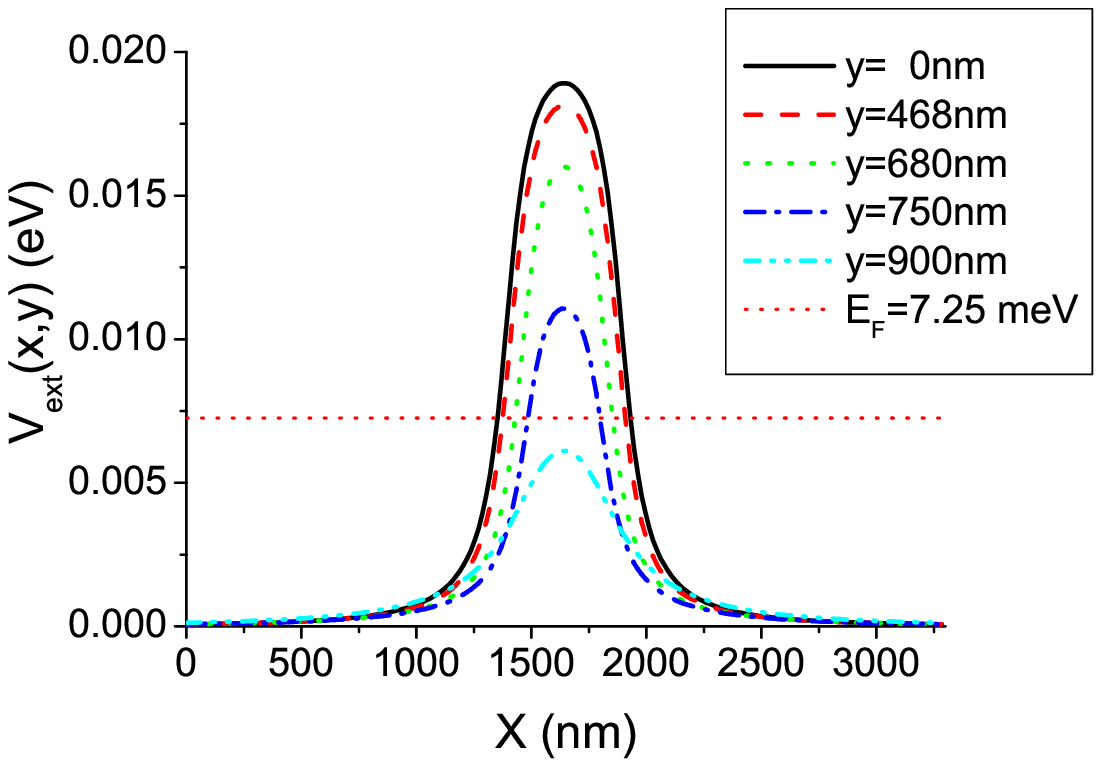}}
\caption{ \label{fig:barepot} The bare confinement potential
generated by the QPC, defined by the polygons shown in figure
\ref{fig:qpcgeo}. The color scale indicates the strength of the
confinement.}
\end{figure}

\begin{figure}
{\centering \includegraphics[width=1.\linewidth]{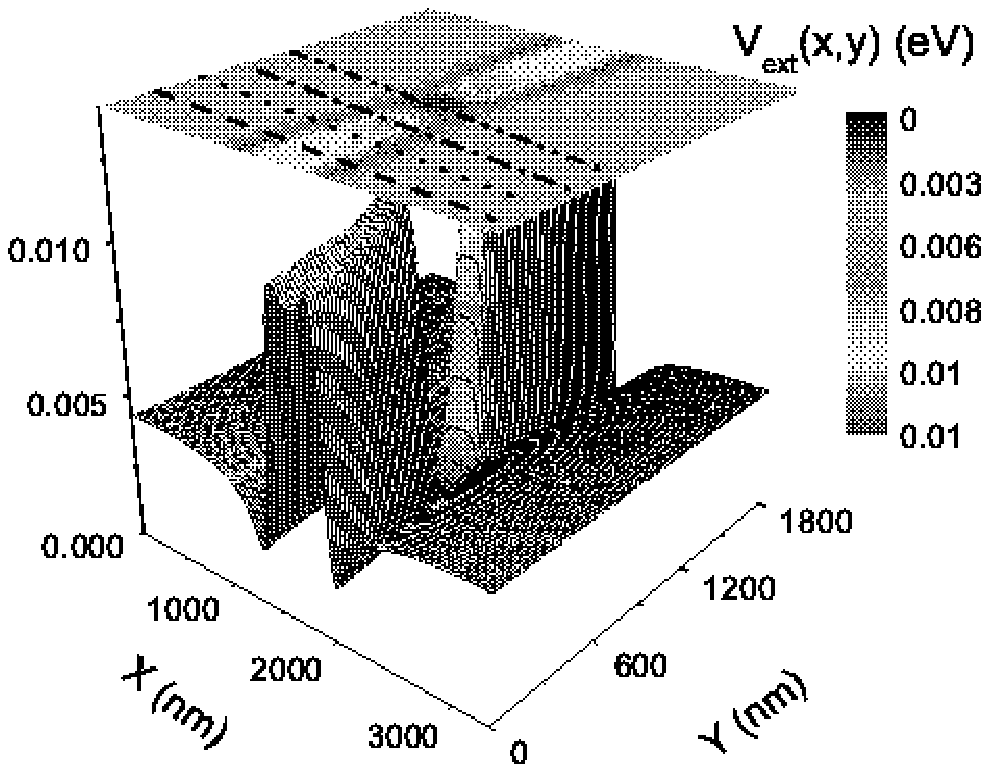}
\includegraphics[width=1.\linewidth]{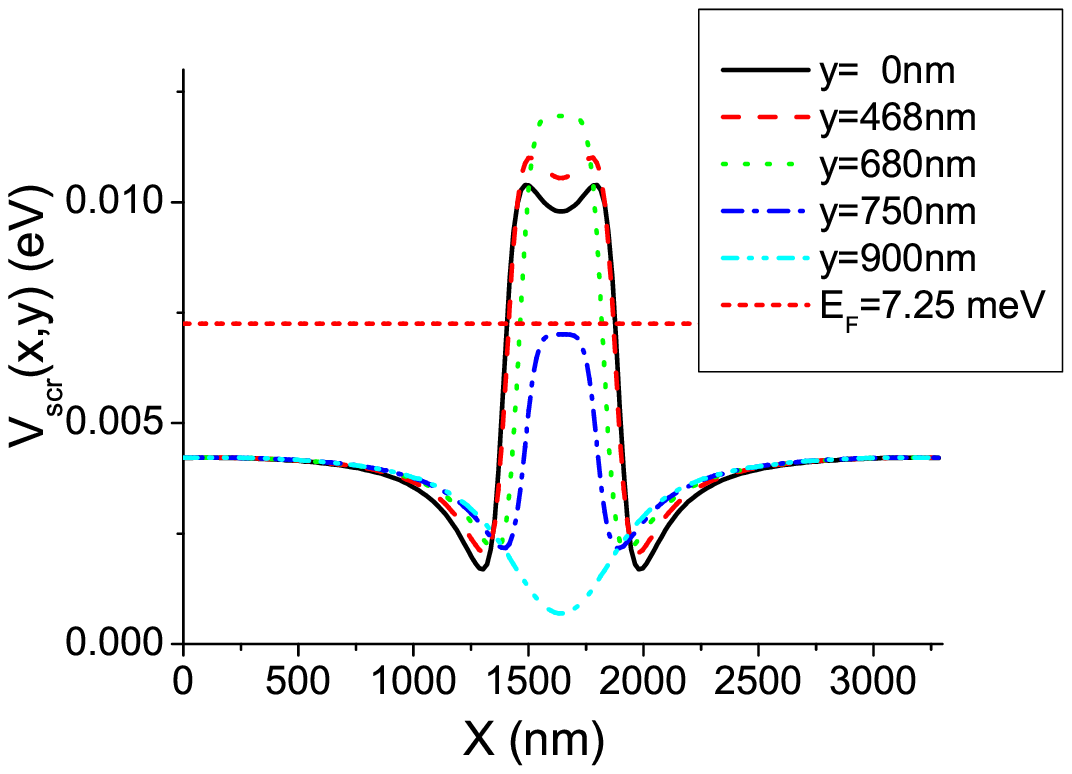}}
\caption{ \label{fig:sqrpotqpc} The screened potential (upper
panel) seen by a 2DES at $85$nm below the surface and some
characteristic cuts along the $x$- axis, together with an
indication of the Fermi level $E_F$ (lower panel). The color scale
represents the strength of the potential, and the cross-sections
are indicated by the same line code as in Fig.~\ref{fig:barepot}.}
\end{figure}

We define the QPC using rectangles and polygons which are shown in
figure \ref{fig:qpcgeo} as red (dark) and white areas.  In figure
\ref{fig:barepot} we show the bare confining potential for the
parameters given above. The electrons are filled up to the Fermi
energy ($E_{F}\sim 7$meV, corresponding to a typical electron
surface density $n_{el}\sim 1.7\times10^{11}$ cm$^{-2}$). Using
such parameters, the full screening calculation to be discussed
below will reveal the electrons to be depleted beneath the QPC,
say at all the dark (blue) regions in Fig.\ref{fig:sqrpotqpc}.

In our numerical simulations, we have mapped the unit cell
containing the QPC of physical dimensions $3.3\mu$m$\times1.8\mu$m
to a matrix of 200 by 200 mesh points in the absence of a magnetic
field and $1.1\mu$m$\times1.8\mu$m to a matrix of 48 by 96 mesh
points in the presence, which allows us to perform numerical
simulations within a reasonable computation time. With regard to
numerical accuracy, we estimate that, for typical electron
densities, the mean electron distance, i.e. the Fermi wavelength,
is larger than 40nm. Hence, the number of mesh points considered
here allows us to calculate the electron density with a good
numerical accuracy. We also performed calculations for finer
meshes and the results do not differ quantitatively (at the
accuracy of line thicknesses), whereas the computational time
grows like the square of the number of the mesh points. We should
also note that due to computation time concerns we had to use a
smaller unit cell in the presence of the magnetic field, which
yields finite size effects close to the boundaries of the sample
(e.g. see Fig~\ref{fig:avfill1}b). The features observed are, in
principle, negligible and they tend to disappear when the unit
cell is taken to be larger and mapped on a larger matrix.

We now discuss the resulting bare and screened potential for a
realistic QPC defined by surface gates, with a tip opening $\Delta
y =$ $300$nm. Figure \ref{fig:barepot} represents the external
potential created by the QPC gate structure at the surface,
calculated in the plane of the 2DES located at $z=85$nm below the
surface, with an applied potential $-0.3$V. In the upper panel we
show a 3D plot and a planar projection, together with four guide
lines, which indicate the location of the cross-sections through
the that are displayed in  the lower panel. The level of the Fermi
energy of the system (to be assumed below) is indicated in the 3D
plot as well. These results have been obtained numerically from
Eq. (\ref{eq:davies}). The barrier is formed by the regions of
elevated potential.

At the first glance one observes that the potential landscape is
smoothly varying. This is purely an effect of the relatively large
distance to the gate, as screening effects have not yet been
included. For the given Fermi energy (obtained from the electron
density in the bulk) and the tip separation, $\Delta y \gtrsim
100$ nm, the number density of electrons inside the QPC opening
fulfills the validity relation of the TFA, i.e.
$n_{el}(\rm{center}) a_{B}^{\star}\gg 1$. At the positions where
the height of the barrier becomes larger than the Fermi energy
(light line in the 3D plot and horizontal dashed line in the lower
panel), the probability to find an electron is zero within the
TFA.

We proceed in our discussion with a comparison of the screened
potential shown in Fig. \ref{fig:sqrpotqpc} to the bare
confinement potential discussed up to now (Fig.
\ref{fig:barepot}). The self-consistent potential is obtained from
the formalism described above for periodic boundary conditions at
zero temperature and zero magnetic field. The electrons are filled
up to the Fermi energy (shown by the gray thick line on the
surface of the color plot and dashed line in the lower panel),
such that no electrons can penetrate classically into the barrier
above those lines. The first observation is that the potential
profile becomes sharper for the screened case and strong
variations are observed in the vicinity of the QPC. These
shoulder-like local maxima near the QPC represent the same feature
seen in the example of the square barrier discussed previously,
and we have pointed out that they stem from $q$-dependent,
non-linear screening. This will become more important when we
consider a magnetic field, since the local "pinning" of the Landau
levels to the Fermi energy in these regions will produce
compressible regions surrounded by incompressible regions.

An interesting feature occurs near the opening of the QPC, namely
a local minimum which is a result of the non-linear screening. We
point out that somewhat similar physics has been found (using
spin-density-functional theory\cite{Hirose03:026804}) to lead to
the formation of a local bound state inside a QPC, which has been
related to the "0.7" anomaly, linking it with Kondo physics. We
believe this feature to be a very important result of the
self-consistent screening calculation, and we note that it may
affect strongly the transport properties of the QPC both in the
presence or absence of a magnetic field. We will discuss the
influence of this local minimum on the formation of the
incompressible strips in section \ref{sec:results}, where we
calculate the density and potential profiles including a strong
perpendicular magnetic field.

It is known from the experiments that the interference pattern and
the transmission properties strongly depend on the structure of
the QPCs, such as the distance of the 2DES from the surface, the
applied gate voltage, the sharpness and the geometry of the edges,
as well as the width of the opening of the QPC. The effect of the
first two parameters can be understood by following the simple
arguments of linear screening as shown for the square gate model:
if the distance from the QPC to the 2DES increases, the potential
profile becomes more and more smooth. The screened potential
changes linearly with the applied gate potential (see
Eq.(\ref{eq:qdep})). The geometric parameters have to be adapted
to the experiment in question. Note that the shape of the QPCs has
already been discussed in the literature (see
Ref.~\onlinecite{Macucci02:39} and references contained therein).
The effect of the size of the QPCs, however, has not been
considered for large $\Delta y$ ($>100$nm), and we believe this to
be an important parameter for the interferometer experiments.

We start our investigation by looking at the opening
of the QPC with increasing tip separation of the metal gates used to define the QPC.
In this section, we work at
zero temperature and magnetic field, with a constant bulk electron
density.

\begin{figure}
{\centering
\includegraphics[width=1.\linewidth]{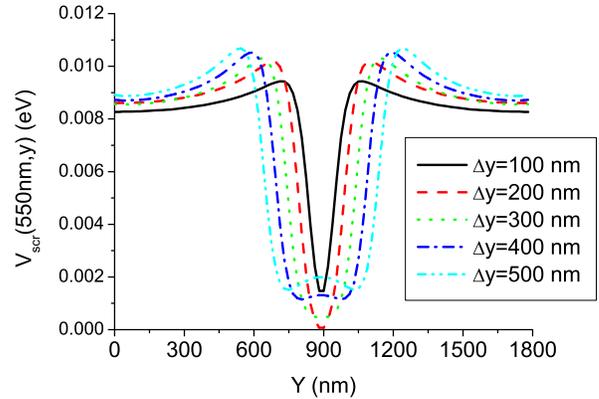}}
\caption{ \label{fig:openqpc} The screened potential at $x=550$nm
for five different tip distances. Note that the $x$ interval used
for the calculation smaller than in the previous figure, since we
concentrate on the bulk structures rather than the edge ones.}
\end{figure}

In figure \ref{fig:openqpc} we depict the self-consistent
potential at the center of the QPC ($y=550$nm), while changing the
tip separation ($\Delta y$) between $100$ and $500$nm. We see that
for the narrowest separation the potential profile looks rather
smooth and a minimum is observed at the center. If we increase
$\Delta y$($\leqslant300$nm) we see that the screening becomes
stronger, leading to more pronounced shoulders on the sides and a
deeper minimum at the center. For even larger separations ($\Delta
y>300$nm) a local maximum starts to develop at the center, since
the electrostatic potential energy is no longer strong enough to
repel the electrons from this region. Basically all the non-linear
features observed result from the competition between the gate
potential, which simply repels the electrons, and the mutual
Coulomb interaction, i.e. the Hartree potential. It is obvious
that for narrower tip separations only a few electrons will remain
inside the QPC opening and therefore TFA type approximations will
not be justified any longer.

Summarizing this section, we have determined the screened
potential profile in a realistic QPC geometry, pointing out
features resulting from non-linear screening. We have observed
that a local extremum occurs at the center of the QPC, and have
traced the dependence on the width $\Delta y$ between the QPC
tips. These features, as mentioned before, become more interesting
if a magnetic field is also taken into account, where they lead to
stronger spatial inhomogeneities in the electron distribution. Our
next step is thus to include a strong quantizing perpendicular
magnetic field and examine the distribution of the incompressible
strips where the imposed external current is
confined\cite{Ahlswede02:165,siddiki2004}.

\subsection{Finite temperature and Magnetic field\label{sec:results}}
\begin{figure}[h] \centering
\includegraphics[width=1.\linewidth,angle=0]{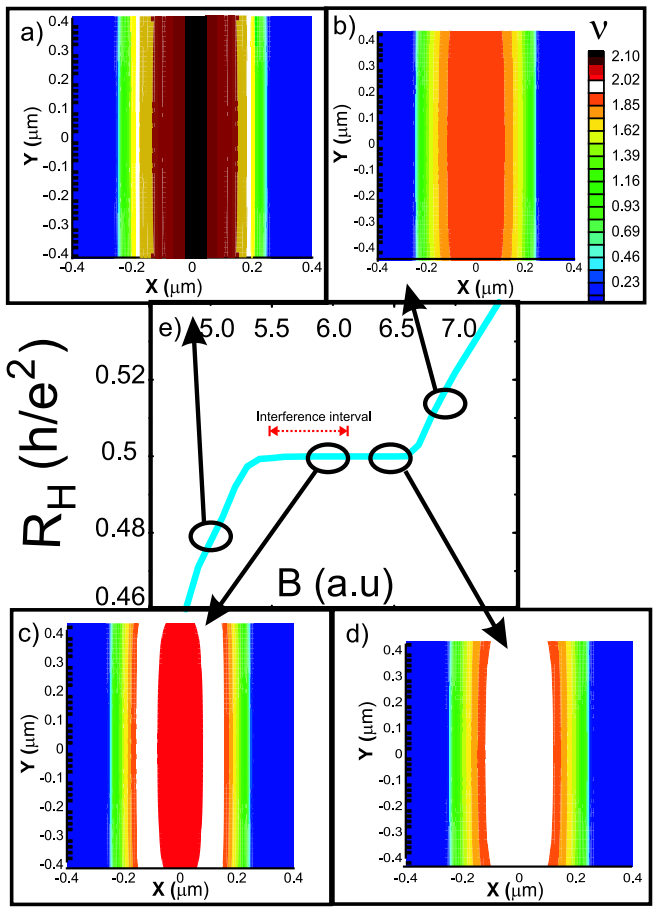}
\caption{\label{fig:sketch1}[a]-[d] Color-coded plot of the local
filling factor versus position $(x,y)$ for a square sample of
width $a_{x}=a_{y}=0.8\, \mu$m; white indicates $\nu(x,y)=2$. The
average density is taken to be $3.0\cdot 10^{11}\,$cm$^{-2}$;
$k_BT/E_F=0.02$. [e] A sketch of the Hall resistance as a function
of magnetic field}
\end{figure}
Once the initial values of the screened potential and the electron
distribution have been obtained for $T=0, B=0$, using the scheme
described above, one can calculate these quantities for finite
field and temperature as follows: replace the zero temperature
Fermi function with the finite temperature one and insert the
\emph{bare} Landau DOS \be \label{landau-dos} D_B(E)=\frac{1}{\pi
l_{b}^{2}}\sum_{n=0}^{\infty}{\delta (E-E_n)}, \quad
E_n=\hbar\omega_{c}(n+1/2)
 \ee into equation (\ref{thomas-fermi}) instead of $D_{0}$. In our
numerical scheme we first start with relatively high temperatures
(i.e. a smooth Fermi function) and then decrease the temperature
slowly until the desired temperature is reached. A Newton-Raphson
method is used for the iteration process and at every iteration
step the electro-chemical potential is checked to be constant.

Before proceeding with the investigation of the QPC geometry at
$B>0$, we would like to make clear the relation between the
quantum Hall plateaus and the existence of the incompressible
strips following the arguments of Siddiki and
Gerhardts~\cite{siddiki2004}. Fig.~\ref{fig:sketch1} presents the
local filling factors of a relatively small Hall bar, together
with an illustrative Hall resistance curve. At the high magnetic
field side ((b), $\nu(0,0)<2$) there are no incompressible strips,
thus the system is out of the Hall plateau. When approaching from
the high $B$ side to the plateau a single incompressible strip at
the center develops. When the width of this strip becomes larger
than the Fermi wavelength, the system is in the quantum Hall state
((d), $\nu(0,0)=2$). If we decrease the field strength further the
center incompressible strip splits into two and moves toward the
edges ((c) $\nu(0,0)>2$). As long as the widths of these strips
are larger or comparable with the Fermi wavelength the system
remains in the plateau. This is the regime in which an
interferometer may be realized. Further decreasing the magnetic
field leads to narrower incompressible strips which finally
disappear if the widths of them become smaller than the average
electron distance. Then the system leaves the quantized plateau.
The distribution of the incompressible strips and the onset of the
plateaus, of course, depends on the disorder
potential~\cite{Siddiki:ijmp} and the physical size of the sample.
However, the experiments considered here are done using narrow and
high mobility structures, thus the above scheme will cover the
experimental parameters.

In this subsection we present some of our results obtained within
the TFA using periodic boundary conditions, considering two
different tip separations, while sweeping the magnetic field.
First we will fix the gate potential to $V_{QPC}=-0.3$V and sweep
the magnetic field for $\Delta y=100$nm, while keeping the
electron number density, i.e. the Fermi energy, constant. Second
we examine the potential profile for $\Delta y=300$nm and comment
on the possible effects on the coherent transport properties.

\begin{figure}[h] \centering
\includegraphics[width=1.\linewidth,angle=0]{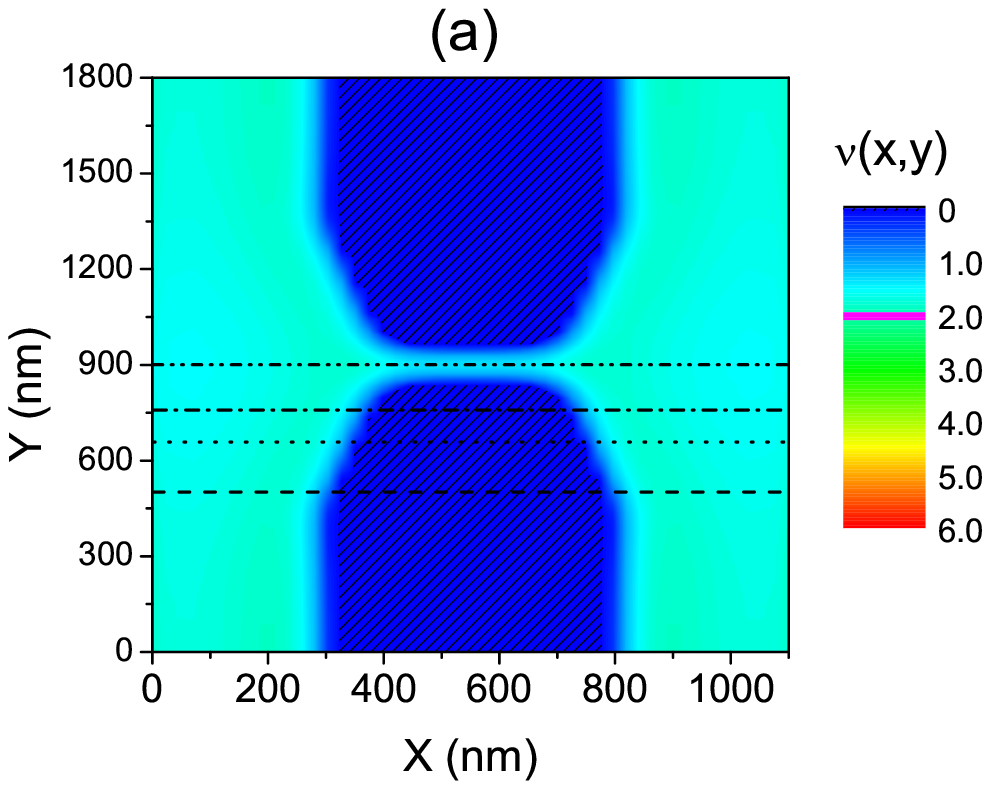}
\includegraphics[width=1.\linewidth,angle=0]{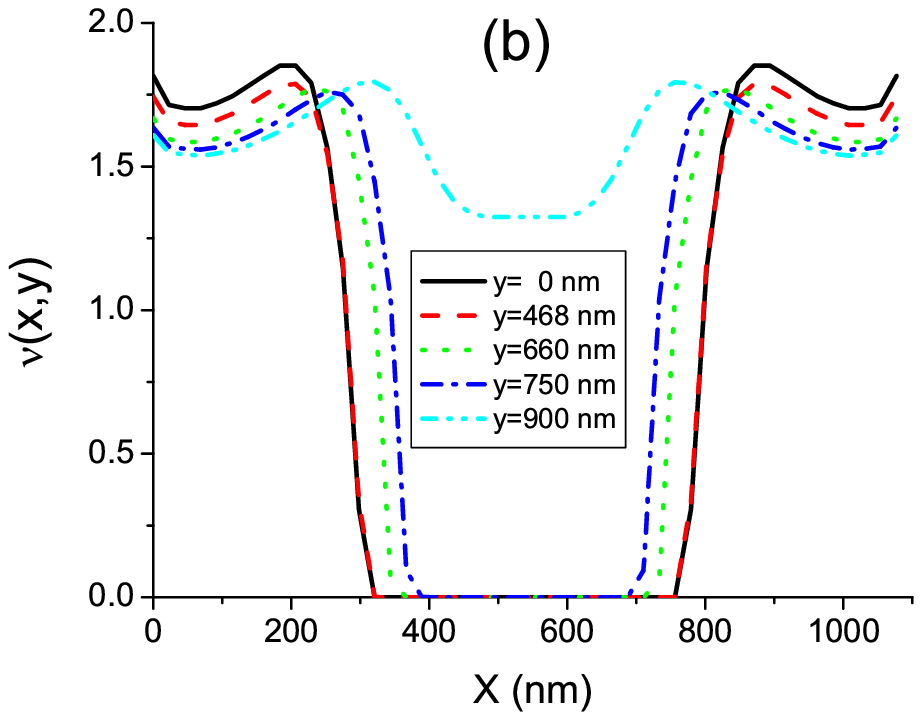}
\caption{\label{fig:avfill1}[a] The top view of the local filling
factor, $\nu (x,y)$, distribution of the 2DES, for average filling
factor one, in the plane located at $z=85$nm below the surface, at
the "default" temperature, $kT/\hbar\omega_{c}=1/50$, which will
be used in all subsequent plot. The color scale depicts the
density of electrons, whereas the dark shaded areas indicate the
electron depleted regions. [b] Side view of the local filling
factor for $y=0$nm(solid line), 468nm (dashed line), 660nm (dotted
line), 750nm (dash-dotted line) and 900nm (dash-dotted-dashed
line) . The horizontal lines in [a] shows the positions of the
cuts in [b], with the same line code. Note that the density has
local minima at large and small x, which are finite size effects
mentioned in the text.}
\end{figure}

In figure \ref{fig:avfill1} we plot the local filling factor (i.e.
the normalized density) distribution of the 2DES projected on the
$xy$-plane, together with the same quantity for some selected
values of $y$,  at average filling factor ($\bar{\nu}$) one. From
the $y=0$nm curve (solid lines) in Fig.\ref{fig:avfill1}[b], one
can see that the electrons beneath the QPC are depleted (shaded,
dark (blue) regions) ($300<x<800$nm), while the electron density
reaches finite values while approaching the opening of the QPC
($y\sim 850$nm). At $\bar{\nu}=1$ one does not observe any
incompressible regions, since the Fermi energy is pinned to the
lowest Landau level. Hence the electron distribution is rather
smooth and the current distribution will just be proportional to
the number of electrons, similar to the Drude approach. For this
case the external potential is screened almost perfectly and the
self-consistent potential is almost flat, thus one can assume that
the corresponding local wave functions are very similar to the
ground state Landau wave functions.

\begin{figure}[h] \centering{
\includegraphics[width=1.\linewidth,angle=0]{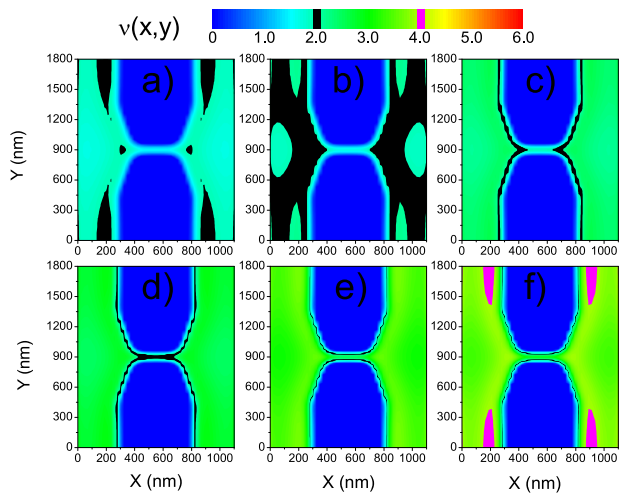}
\caption{\label{fig:avfill2345}The local filling factor
distribution for different average filling factors ($\bar{\nu}$),
which is defined by the number of the electrons in the unit cell.
[a] $\bar{\nu}= 1.1$ [b] $1.2$ [c] $1.4$ [d] $1.6$ [e] $1.8$ [f]
$2.2$. The color scale depicts the local electron concentration,
whereas the abrupt colors indicate the even-integer filling
factors, i.e. incompressible strips, (black for $\nu(x,y)=2$,
magenta for $\nu(x,y)=4$.}}
\end{figure}

The first incompressible region occurs when the Fermi energy falls
in the gap between two low-lying Landau levels. Then the electrons
exhibit a constant density and thus cannot screen the external
potential. In figure (\ref{fig:avfill2345})a, we show the electron
distribution for $\bar{\nu}=1.1$. The black regions denote a local
density corresponding to filling factor $\nu=2$, which does not
percolate from the left side of the sample (which we might
identify with the source) to the right side (drain). Here one can
see well developed incompressible puddles, at the regions
$150$nm$<x<250$nm, $0$nm$<y<450$nm (and four other symmetric ones)
and two smaller puddles at the entrance of the QPC. These
structures will remain unchanged even if one considers a larger
unit cell, since they manifest the $q$- dependency, i.e. the rapid
oscillations of the Fourier transform of the confining potential
of the QPC.

In these regions the self-consistent potential exhibits a finite
slope. Accordingly the wave functions will be shifted and
squeezed, i.e. they are now superpositions of a few high order
Landau wave functions with re-normalized center coordinates. This
behavior has been shown\cite{Suzuki93:2986,siddiki2004} for the
translationally  invariant model. Here we did not include the
finite extent of the wave functions, here to avoid lengthy
numerical calculations.

The incompressible regions shift their positions on the $xy$ plane
depending on the strength and the profile of the confining
potential. In figure (\ref{fig:avfill2345})b we show the filling
factor distribution where the bulk filling factor is almost two.
We see that four incompressible strips are formed near the QPC.
However the QPC opening remains in a compressible state, with
local filling factor less than two, where we expect that the
self-consistent potential is essentially flat. Further increasing
the average filling factor, we observe that the bulk becomes
completely compressible and two incompressible strips are formed
near the QPC which percolate from bottom to top, creating a
potential barrier with a height of $\hbar \omega_{c}$, see Fig.
(\ref{fig:avfill2345}c). For even higher filling factors, they
merge at the center of the QPC (Fig.~\ref{fig:avfill2345}d). In
that case, the potential within the QPC will then no longer be
flat, due to poor screening. We should also note that for a small
width $\Delta y$ of the QPC opening, merging of the incompressible
strips will happen only in a very narrow $B$ interval, and a
quantitative evaluation within our TFA can not be always
satisfactory, as the number of electrons inside the QPC becomes
too low. Further decreasing the field strength (increasing the
average filling factor) results in two separate incompressible
strips winding around the opposite gates making up the QPC, as
shown in Fig. \ref{fig:avfill2345}e. Thus, dissipationless
transport through the QPC, with a quantized conductance, becomes
possible. At the lowest field values considered in this figure, we
see that the innermost incompressible strips (with $\nu=2$) become
smaller than the Fermi wavelength and thus they essentially
disappear  and no longer affect the transport properties. This
point has been discussed in detail in a recent work by Siddiki et
al.\cite{siddiki2004}. The scheme described above now starts to
repeat, but with incompressible strips having a local filling
factor of 4.

\begin{figure}[h] \centering{
\includegraphics[width=1.\linewidth,angle=0]{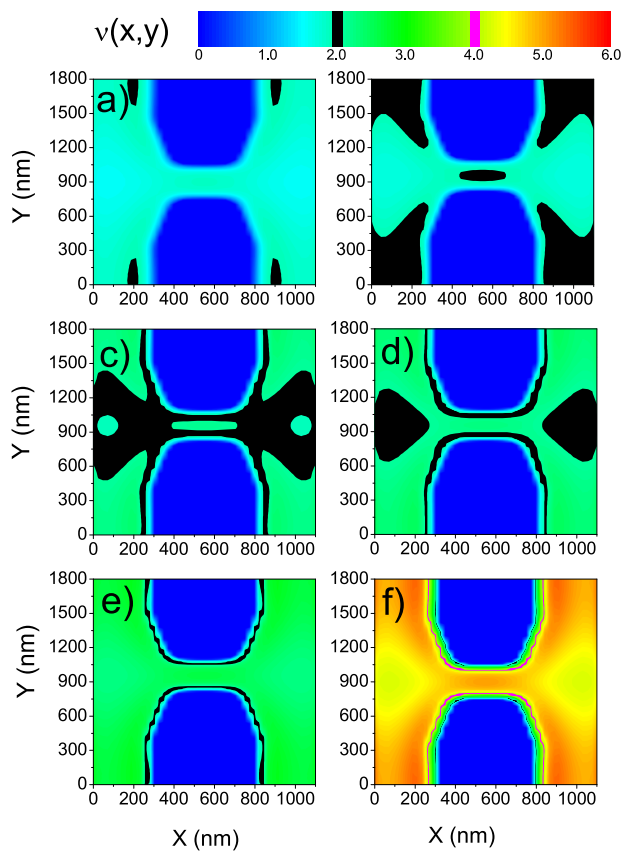}
\caption{\label{fig:avfill23456}The local filling factor
distribution for different average filling factors ($\bar{\nu}$),
for a tip separation $\delta y=300$ nm. Note that, the number of
electrons in the unit cell is changed, since the depleted areas
are larger than of the previous figure. [a] $\bar{\nu}= 1.14$ [b]
$1.2$ [c] $1.34$ [d] $1.4$, [e] 1.6 and [g] 3.1. The color scale
depicts the local electron concentration, whereas the
high-contrast color regions indicate the even-integer filling
factors, i.e. incompressible strips, (black for $\nu(x,y)=2$,
magenta for $\nu(x,y)=4$). The calculations are done at
$k_{B}T/\hbar\omega_{c}=1/50$ for an average electron density
$1.7\times 10^{-11}cm^{-2}$.}}
\end{figure}

We now discuss the effects of increasing the separation parameter,
which we choose to be $\Delta y=300$nm in figure
(\ref{fig:avfill23456}). At the strongest magnetic field
(\ref{fig:avfill23456}a), only very small regions are
incompressible and the electron distribution is similar to
Fig.\ref{fig:avfill2345}a, where the incompressible regions result
from local unpinning of the Fermi energy from the lowest Landau
level due to $q-$dependent screening, i.e the shoulder-like
variation of the potential near the QPC discussed earlier. By
decreasing $B$, an interesting structure is observed at the center
of the QPC: an incompressible island. In figure
\ref{fig:avfill23456}b, we have tuned the magnetic field such that
the bulk of the 2DES is incompressible, meanwhile the entrance to
the QPC remains compressible. The strong variation of the
self-consistent potential at the center of the QPC can generate a
pronounced effect on the current passing through the QPC (see
figure \ref{fig:centerpot} and the related text). For a lower
magnetic field strength the distribution of the incompressible
region is just the opposite (c). Now we see a large compressible
puddle at the center, surrounded by incompressible regions, which
can percolate from source to drain. Coherent, dissipationless
transport can be expected in this case. Further decreasing the
magnetic field we observe that the structure is smeared out and
the tip region becomes compressible, nevertheless there are two
large incompressible regions close to the entrance of the opening.
The two incompressible strips wind around the gates, as shown in
Fig. \ref{fig:avfill23456}d. Finally, a scheme similar to that
observed earlier in figure \ref{fig:avfill2345}d-e is also seen
now, for $\Delta y =300$nm.

Another remark which we would like to make concerns the edge
profile of the sample itself and of the QPC. It has been shown
both experimentally\cite{Grayson05:016805} and
theoretically\cite{Wulf88:4218,siddiki2004} that for an (almost)
infinite potential barrier at the edges of the sample, the
Chklovskii [\onlinecite{Chklovskii92:4026}] edge state picture
breaks down, i.e. no incompressible strips near the edge can be
observed. Meanwhile for smoothly varying edge potential profiles
many incompressible strips are present, if the bulk filling factor
is larger than two (for spinless electrons: four). We believe
that, within the MZI setup both of these edge potential profiles
might co-exist. At the edge regions of the sample, where lateral
confinement is defined by physical etching, the potential profile
differs from of the one generated by the top gates, due to
different separation thicknesses and also lateral surface charges
generated by etching. In principle gate and etching defined edges
impose different boundary conditions, and the effects on screening
at a 2DES have been discussed before\cite{Siddiki03:125315}. These
two profiles will certainly affect the group velocity, since the
slope of the potential depends on the (lateral) boundary
conditions. Following the arguments of
Ref.[\onlinecite{Guven03:115327,siddiki2004}], which essentially
predict that the dissipative current is confined to the
incompressible strips, the widths of these strips will also define
the slope, hence the velocity of the electrons will be determined
by the edge profile. The velocity of the edge electrons were
investigated experimentally\cite{Karmakar05:282} and the magnetic
field dependency was reported as $B^{3/2}$. There it was noted
that a self-consistent treatment is necessary to understand their
findings, which we would like to discuss in a future publication.

The important features to note in these results are (i) in
general, electron-electron interactions have a remarkable effect,
leading to the formation of a local extremum in the potential at
the center of the QPC, which even at low electron densities seems
to be well described by the TFA; (ii) the narrow
compressible/incompressible strips formed near the QPC are a
direct consequence of the $q-$dependent screening.

\subsection{Comments on coherent transport}

A complete calculation of coherent transport requires a deeper
analysis of the wave functions and is beyond the scope of this
work, which has been devoted to self-consistent realistic
calculations of the potential and density profiles. In principle,
one can follow the arguments of the well developed recursive
Green's function technique\cite{Solsl89:3892} in the absence of
magnetic field and the method developed recently even in the
presence of a strong field\cite{Rotter03:165302}.

\begin{figure}[h] \centering{
\includegraphics[width=1.\linewidth,angle=0]{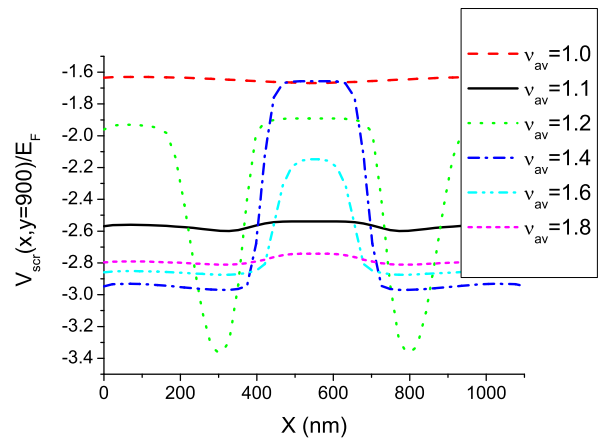}
\caption{\label{fig:centerpot}The self-consistent potential
profile across the QPC, plotted for characteristic
values of the average filling factor. Calculations are done at the
default temperature and electron density.}}
\end{figure}

Instead we would like to examine the potential distribution across
the QPC and comment on the possible effects of interaction on the
wave functions, and thereby (indirectly) on transport. In figure
\ref{fig:centerpot}, we depict the potential profile across the
QPC for the parameters used to obtain figure \ref{fig:avfill2345}.
As expected for $\bar{\nu}=1.0$ (dashed (red) line) the 2DES is
"quasi" metallic, hence the external potential is perfectly
screened, and the wave functions are left almost unchanged. The
two incompressible islands seen at the entrance of the QPC in
Fig.\ref{fig:avfill2345}a lead to a minor variation of the
screened potential at $x=300$nm and $x=800$nm, depicted by the
solid (black) line for $\bar{\nu}=1.1$. A drastic change is
observed when the bulk becomes incompressible ($\bar{\nu}=1.2$)
and the opening remains compressible: Now the 2DES cannot screen
the external potential near the openings of the QPC, where we see
a strong variation. The strong perpendicular magnetic field
changes the potential profile near the QPC via forming
incompressible strips, and local minima are observed at the
entrance and the exit. In these regions the electrons are strongly
localized and the wave functions are squeezed. The situation is
rather the opposite for $\bar{\nu}=1.4$, where two incompressible
strips located near the QPC, formed due to $q$-dependent
screening, merge at the opening. One observes a barrier with the
height of $\hbar\omega_c$, which essentially is a direct
consequence of the incompressible strip at the center and
electrons have to overcome this barrier. Further decreasing the
magnetic field smears out the barrier gradually, until the system
becomes completely compressible and we are back in the case of
figure \ref{fig:centerpot}a (also with regard to the transport
properties).

\section{\label{sec:summary}Summary}

The study that motivated the present authors was the Quantum Hall
effect based Mach-Zender
interferometer\cite{Heiblum03:415,Neder06:016804}. There are
puzzles in the experiment for which it is not obvious (at least to
us and some others) how they could be explained using scattering
theory. Therefore we probably need to take into account
interactions more seriously, including correlation effects in
non-equilibrium transport that will not be part of scattering
theory. A first step towards that goal then is the self-consistent
mean-field equilibrium calculation which we have done.

At the moment we do not have to offer a non-equilibrium transport
theory based on these equilibrium calculations. These effects may
include decoherence due to potential fluctuations brought about by
electron-electron or electron-phonon interactions (together with
other noise sources).  A more detailed understanding of
electron-electron interactions in this setup, as well as of those
features of the interferometer that are specific to the physics of
the Quantum Hall effect, hinges on an analysis of the
self-consistent static potential landscape near the QPCs, which
represent the most crucial components of the setup.

Therefore, in this work, we have taken into account the
electron-electron interaction within the TFA, considering
realistic geometries of QPCs, calculating the self-consistent
potential and electronic density profiles, and commenting on
possible effects on transport.

The outcome of our model calculations can be summarized as
follows: (i) We have obtained the electrostatic potential profile
for the QPC geometries used in the experiments by solving the
Laplace equation semi-analytically. (ii) We have demonstrated for
a simple square well barrier that the screened potential in a
2DES, even in the absence of a magnetic field, is strongly
dependent on the initial potential profile and on the distance
between gates and 2DES. (iii) The screened potential has been
calculated within the TFA for a QPC at vanishing field, where we
have observed two interesting features: a local extremum at the
center of the QPC and strong shoulder-like variations near the
QPC. (iv) In the presence of a magnetic field, the formation and
the evolution of the incompressible regions has been examined and
three different cases have been observed: (a) the system is
completely compressible. (b) An incompressible region and/or
strip, which does not percolate from source to drain, generates a
local extremum at the entrance/exit of the QPC (c) The center of
the QPC becomes incompressible, with or without a compressible
island, hence the incompressible strip percolates from source to
drain. We note that the local minimum found at the center of the
QPC for certain tip separations, being a clear interaction effect,
coincides with the findings of Hirose \emph{et al}
[\onlinecite{Hirose03:026804}].

This work was partially supported by the German Israeli Project
Cooperation (DIP) and by the SFB 631. One of the authors (A.S)
would like to acknowledge R.R. Gerhardts, for his supervision,
support and discussions, J. von Delft for offering the opportunity
to work in his distinguished group and S. Ludwig for discussions
on the experimental realization of QPC structures.

\end{document}